\documentclass[12pt]{article}
\usepackage{amsmath,amssymb,amsthm}
\usepackage{mathtools}
\usepackage{bbm}
\usepackage{graphicx}
\usepackage{epstopdf}
\usepackage{psfrag}

 \DeclareMathOperator{\tr}{Tr}

\newcommand{\bi}[1]{\boldsymbol{#1}}

\newcommand{\s}[1]{\mathsf{#1}}

\theoremstyle{definition}

\begin{document}
\nocite{*}
\setlength{\parindent}{0pt}
\setlength{\parskip}{6pt plus 2pt}

\title{Monotonicity of the von~Neumann entropy expressed as a function of R\'enyi entropies}
\date{October 22 2013}

\maketitle

\begin{center}
Mark~Fannes\footnote{{\bf Email:} mark.fannes@fys.kuleuven.be}\\[6pt]
Instituut voor Theoretische Fysica \\
KU~Leuven, Belgium \\[6pt]
\end{center}

\begin{abstract}
\noindent
The von~Neumann entropy of a density matrix of dimension $d$, expressed in terms of the first $d-1$ integer order R\'enyi entropies, is monotonically increasing in R\'enyi entropies of even order and decreasing in those of odd order.


\end{abstract}

This paper is about the monotonicity of the von~Neumann entropy expressed as a function of integer order R\'enyi entropies. As entropies are unitarily invariant quantities associated with a single density matrix $\rho$, we can restrict our attention to diagonal density matrices, i.e.\ probability vectors. The integer R\'enyi entropy of order $q = 2, 3, \ldots$ is usually defined as~\cite{ren,ben,ohy}
\begin{equation}
\s S_q(\rho) := - \frac{1}{q-1}\, \log\Bigl( \tr \rho^q \Bigr).
\label{renyi}
\end{equation}
and the von~Neumann entropy equals formally the R\'enyi entropy of order 1
\begin{equation}
\s S(\rho) := - \tr \rho \log\rho, 
\end{equation}
where $0 \log 0 := 0$.

Entropic quantities are relevant for translation-invariant many particle systems where the entropies of physically relevant states are typically proportional to the number of particles. In such a situation the R\'enyi and von~Neumann entropies per particle $\s s_q$ and $\s s$ are important quantities. It is known that the average R\'enyi entropies don't always exist and that they lack in general good continuity or convexity properties. Still, for particular subclasses of states, e.g.\ states with good cluster properties, the R\'enyi densities are meaningful. One of their major advantages is that the low order densities can sometimes be computed rather explicitly using multiple independent copies of the system, this is the replica trick, see e.g.~\cite{alm}. `Taking the limit $q \to 0$' is a widely used approach in statistical physics. The general question of relating R\'enyi and von~Neumann entropies is therefore important~\cite{zyc,ber}. A number of interesting bounds obtained in finite dimensions don't survive the thermodynamic limit and there are quite few general relations available between the densities, provided they exist. The aim of this note is to prove in $d$ dimensions, a monotonicity property of $\s S$, expressed as a function of $\s S_2$, $\s S_3,\ldots$, $\s S_d$.

We first recall some basic notions, see~\cite{meh} for background material. Consider sequences $\bi \lambda = (\lambda_0, \lambda_1, \ldots)$ of complex numbers with only a finite number of entries different from 0. The elementary symmetric polynomials $e_k$, $k = 0, 1, 2,\ldots$ are defined as follows
\begin{equation}
e_0 = 1,\enskip e_1 = \sum_j \lambda_j,\enskip e_2 = \sum_{j_1 < j_2} \lambda_{j_1} \lambda_{j_2},\enskip \cdots
\label{e}
\end{equation}   
The entries of $\bi \lambda$ are non-negative if and only if all $e_k$ are non-negative. For the remainder of this note we restrict ourselves to probability vectors $\bi \lambda$ of length $d$ with non-increasing entries. It is then well-known that the symmetric polynomials $e_2$, $e_3,\ldots$, $e_d$ completely determine $\bi \lambda$.

We also need the power sums of sequences of length $d$   
\begin{equation}
r_0 = d,\enskip r_1 = \sum_j \lambda_j,\enskip r_2 = \sum_j \lambda^2_j,\ \cdots
\label{r}
\end{equation}   
Again, the power sums $r_2$, $r_3,\ldots$, $r_d$ fully determine $\bi \lambda$.

The powers sums can be expressed as polynomials in elementary symmetric invariants and vice versa:
\begin{equation*}
2e_2 = 1 - r_2,\enskip 6e_3 = 1 - 3r_2 + 2r_3,\enskip 24e_4 = 1 -6r_2 + 3r^2_2 + 8r_3 -6r_4,\enskip \cdots 
\end{equation*}
and
\begin{equation*}
r_2 = 1 - 2e_2,\enskip r_3 = 1 - 3e_2 + 3e_3,\enskip r_4 = 1 -4e_2 + 2e^2_2 + 4e_3 -4e_4, \enskip \cdots 
\end{equation*}

Also the entropies $\s S_q$ and $\s S$ can both be expressed, either as functions of $e_2$, $e_3,\ldots$, $e_d$ or of $r_2$, $r_3,\ldots$, $r_d$. It was shown in~\cite{mit} that $\s S$ is an increasing function of the elementary symmetric invariants.

It is our aim to show that $\s S$ is decreasing in the power sums of even order and increasing in these of odd order. This is, in view of~(\ref{renyi}), equivalent to
\begin{equation}
\frac{\partial \s S}{\partial \s S_q} \ge 0 \enskip\text{for $q$ even and}\enskip
\frac{\partial \s S}{\partial \s S_q} \le 0 \enskip\text{for $q$ odd.}
\label{res}
\end{equation}  

We first provide an elementary proof of $\partial \s S / \partial e_k \ge 0$, based on the integral representation
\begin{equation}
-x \log x = 1 - x - \int_0^\infty \!dt\, \Bigl\{ \log(t+1) - \log(t+x) - \frac{1-x}{t+1} \Bigr\},\enskip x \ge 0.
\end{equation}
Applying this to a density matrix of dimension $d$ yields
\begin{equation}
\s S = d - 1 - \int_0^\infty \!dt\, \Bigl\{ d \log(t+1) - \log\bigl( \det(t+\rho) \bigr) - \frac{d-1}{t+1} \Bigr\}.
\label{1}
\end{equation}
Using the generating function for the elementary symmetric invariants
\begin{equation}
\det(t + \rho) = \sum_{j=0}^d t^{d-j}\, e_j
\label{2}
\end{equation}
we obtain the monotonicity property
\begin{equation}
\frac{\partial \s S}{\partial e_k} = \int_0^\infty \!dt\, \frac{t^{d-k}}{\sum_{j=0}^d t^{d-j}\, e_j} \ge 0\enskip \text{for $k = 2, 3, \ldots, d$.}
\label{mj}
\end{equation}

Next, we express the elementary symmetric invariants $e_2$, $e_3,\ldots$, $e_d$ in function of the first $d-1$ power sums $r_2$, $r_3,\ldots$, $r_d$ and show that for $k = 2,3,\ldots,d$
\begin{equation}
\frac{\partial e_k}{\partial r_\ell} =
\begin{cases}
(-1)^{\ell+1}\, \frac{1}{\ell}\, e_{k-\ell} &\text{for $\ell = 2, 3, \ldots, k$} \\[6pt]
0 &\text{for $\ell = k+1, k+2, \ldots, d$.}
\end{cases}
\label{er}
\end{equation}
In matrix form this relation reads
\begin{equation}
\frac{\partial(e_2, e_3, \ldots)}{\partial(r_2, r_3, \ldots)} =
\begin{pmatrix*}[r]
-\tfrac{1}{2} &0 &0 &\cdots \\[3pt]
-\tfrac{1}{2} &\tfrac{1}{3} &0 & \\[3pt]
-\tfrac{1}{2}\, e_2 &\tfrac{1}{3} &-\tfrac{1}{4} & \\[3pt]
-\tfrac{1}{2}\, e_3 &\tfrac{1}{3}\, e_2 &-\tfrac{1}{4} & & \\[3pt]
-\tfrac{1}{2}\, e_4 &\tfrac{1}{3}\, e_3 &-\tfrac{1}{4}\, e_2 & \\[3pt]
\vdots &&&\ddots 
\end{pmatrix*}
\end{equation}
The proof of~(\ref{er}) is actually quite simple if we start from the identity
\begin{equation}
\sum_{k=0}^d (-1)^k r_k e_{d-k} = 0
\label{id}
\end{equation}
that can be verified by direct inspection after plugging in the definitions of $e$ and $r$, see~(\ref{e}) and~(\ref{r}). Partially differentiating~(\ref{id}) with respect to $r_j$ and observing that $\partial e_k / \partial r_\ell = 0$ for $\ell > k$ yields~(\ref{er}).

Combining~(\ref{renyi}), (\ref{mj}), and~(\ref{er}) we obtain
\begin{equation}
\frac{\partial \s S}{\partial \s S_q} \ge 0\enskip \text{for $q$ even and } \frac{\partial \s S}{\partial \s S_q} \le 0\enskip \text{for $q$ odd.}
\label{mono}
\end{equation} 

Actually a slightly more involved computation shows that also
\begin{equation}
\frac{\partial r_k}{\partial e_\ell} \le 0\enskip \text{for $\ell$ even and } \frac{\partial r_k}{\partial e_\ell} \ge 0\enskip \text{for $\ell$ odd.}
\label{mono2}
\end{equation} 
This implies that~(\ref{mono}) is actually equivalent to~(\ref{mj}). In principle, (\ref{mono}) is better adapted to a situation where a thermodynamic limit has to be taken as the average elementary symmetric invariants don't make sense in such a situation while monotonicity is preserved.

In~\cite{fan} an explicit reconstruction of the von~Neumann average entropy in terms of average R\'enyi entropies was obtained for quasi-free Fermionic states. Successive approximations of $\s s$ by linear combinations of $\s s_q$ exhibit an alternating sign behaviour consistent with~(\ref{res}).


\begin{thebibliography}{99}

\bibitem{ren}
A.~R\'enyi,
On measures of entropy and information,
\emph{Fourth Berkeley Symp.\ Math.\ Stat.\ Prob.},
Mathematical Institute, Hungarian Academy of Sciences, vol.~I, 547 (1961)

\bibitem{ben}
I.~Bengtsson and K.~{\.Z}yczkowski,
\emph{Geometry of Quantum States: An Introduction to Quantum Entanglement},
Cambridge University Press, Cambridge (2006)

\bibitem{ohy}
M.~Ohya and D.~Petz,
\emph{Quantum Entropy and Its Use},
Springer, Berlin Heidelberg New York (1993)

\bibitem{alm}
J.R.L.~de Almeida and D.J.~Thouless,
Stability of the Sherrington-Kirkpatrick solution of a spin glass model,
\emph{Journal of Physics A: Mathematical and General} \textbf{11}, 983 (1978)

\bibitem{zyc}
K.~\.Zyczkowski,
R\'enyi extrapolation of Shannon entropy,
\emph{Open Systems \& Information Dynamics} \textbf{10}, 297 (2003)

\bibitem{ber}
D.W.~Berry and D.C.~Sanders,
Bounds on general entropy measures,
\emph{Journal of Physics A: Mathematical and General} \textbf{36}, 12255 (2003)

\bibitem{meh}
M.L.~Mehta,
\emph{Matrix Theory, Selected Topics and Useful Results},
Hindustan Publishing Corporation (1989)

\bibitem{mit}
G.~Mitchison and R.~Jozsa, 
Towards a geometrical interpretation of quantum-information compression,
\emph{Phys.\ Rev.\ A} \textbf{69}, 032304 (2004)

\bibitem{fan}
M.~Fannes and N.~Van Ryn,
Connecting the von Neumann and R\'enyi entropies for fermions
\emph{J.\ Phys.\ A: Math.\ Theor.} \textbf{45}, 385003 (2012)

\end{thebibliography}
\end{document}